\documentclass[preprint,12pt,authoryear]{elsarticle}
\usepackage{lineno,hyperref}
\usepackage{tabu}
\usepackage{float}
\usepackage[section]{placeins}
\usepackage{rotating}
\usepackage{booktabs}
\usepackage{threeparttable}
\usepackage{adjustbox}
\usepackage{graphicx}  %Required
\usepackage{booktabs} % For formal tables
\usepackage{subfigure}
\usepackage{multirow}
\usepackage{epstopdf}
\usepackage{amsfonts}

\usepackage{algpseudocode}
\usepackage{amsmath}
\usepackage{natbib}
\usepackage{todonotes}

\modulolinenumbers[1]

\usepackage{amssymb}
\journal{Journal of Informetrics}

\begin{document}

\begin{frontmatter}

%% Title, authors and addresses

%% use the tnoteref command within \title for footnotes;
%% use the tnotetext command for theassociated footnote;
%% use the fnref command within \author or \address for footnotes;
%% use the fntext command for theassociated footnote;
%% use the corref command within \author for corresponding author footnotes;
%% use the cortext command for theassociated footnote;
%% use the ead command for the email address,
%% and the form \ead[url] for the home page:
%% \title{Title\tnoteref{label1}}
%% \tnotetext[label1]{}
%% \author{Name\corref{cor1}\fnref{label2}}
%% \ead{email address}
%% \ead[url]{home page}
%% \fntext[label2]{}
%% \cortext[cor1]{}
%% \address{Address\fnref{label3}}
%% \fntext[label3]{}

\title{Predicting the Citations of Scholarly Paper}
\author{Xiaomei Bai$^a$}
\author{Fuli Zhang$^b$*}
\cortext[cor1]{Corresponding author}
\ead{zfuli@outlook.com}
\author{Ivan Lee$^c$}
\address{$^a$ Computing Center, Anshan Normal University, Anshan, China}
\address{$^b$ Library, Anshan Normal University, Anshan, China}
\address{$^c$ School of Information Technology and Mathematical Sciences, University of South Australia, Australia}

\begin{abstract}
Citation prediction of scholarly papers is of great significance in guiding funding allocations, recruitment decisions, and rewards. However, little is known about how citation patterns evolve over time. By exploring the inherent involution property in scholarly paper citation, we introduce the Paper Potential Index (PPI) model based on four factors: inherent quality of scholarly paper, scholarly paper impact decaying over time, early citations, and early citers' impact. In addition, by analyzing factors that drive citation growth, we propose a multi-feature model for impact prediction. Experimental results demonstrate that the two models improve the accuracy in predicting scholarly paper citations. Compared to the multi-feature model, the PPI model yields superior predictive performance in terms of range-normalized RMSE. The PPI model better interprets the changes in citation, without the need to adjust parameters. Compared to the PPI model, the multi-feature model performs better prediction in terms of Mean Absolute Percentage Error and Accuracy; however, their predictive performance is more dependent on the parameter adjustment.
\end{abstract}

\begin{keyword}
\textbf{Scholarly Paper \sep Paper Potential Index \sep Multi-feature Model}
\end{keyword}

\end{frontmatter}

%% \linenumbers

%% main text
\section{Introduction}
There is an increasing interest in understanding the citation dynamics of scholarly paper and the evolution in science~\citep{Xia2017Big}. So far, studies in this field have primarily been focused on success of science~\citep{Feng2016Bibliographic,Bai2016Identifying,Cao2016A,Fiala2018PageRank,Zhang2017Exploring}, academic collaboration networks~\citep{Panagopoulos2017Detecting}, team science~\citep{Ledford2015Team} and scientific impact prediction~\citep{Bai2017An}. While citation serves as a popular indicator for measuring the research outcome, it is often required to estimate the future impact as well. For instance, research impact prediction helps in effective allocation of research funds~\citep{clauset2017data}. An important challenge in scientific impact prediction is to characterize the change in citations over time, and it is important to identify the factors that affect citations of scholarly papers.

Previous studies have mainly focused on predicting the citations or analyzing future citation distributions. Some studies utilize machine learning algorithms such as Gradient Boosting Decision Tree~\citep{sandulescu2016predicting}, Support Vector Machine~\citep{Adankon2010Support}, and XGBoost~\citep{Chen2016XGBoost}. To train the validity of the predictive models, crucial features have been identified for citation prediction, including early citations, journal impact factor, authors' authority, journal reputation, topic of scholarly paper, and age~\citep{petersen2014reputation,Sarig2014Predicting,yu2014citation}. Some citation prediction studies have applied generative model to reflect the observation that older papers typically attracted higher citations~\citep{Newman2008The}, or to address some citation patterns that come with an initial period of growth followed by a gradual declined over time~\citep{Wang2008Measuring,wang2013quantifying}.
More recently, \cite{xiao2016modeling} proposed a point process model to predict the long-term impact of individual publications based on early citations. Furthermore, \cite{Singh2017Understanding} has found that early influential citers negatively affected long-term scientific impact, possibly due to attention stealing, whereas non-influential early citers positively affected long-term scientific impact.

Inspired by the prior work \cite{Wang2008Measuring,wang2013quantifying,xiao2016modeling,Singh2017Understanding}, we model the Paper Potential Index (PPI) by considering the following factors: inherent quality of scholarly paper,
scholarly paper impact decaying over time, early citations, and early citers' impact. The PPI predictive model combines these factors and expands the Hawkes process, and it mainly depends on the inherent involution mechanism of paper citations with the following three properties:
(1) Paper citation declines along with the decay of paper novelty over time;
(2) The early citer's impact can increase scholarly paper impact in the predictive model;
(3) Early citations help retaining long term citations.

In addition, we propose a multi-feature predictive model, which considers author-based features, journal-based features, and citations feature. We compare the prediction results of the two models in terms of mean absolute error, root mean squared error, range-normalized RMSE, mean absolute percentage error and accuracy.

Main contributions of this paper include: (1) Introduction of PPI which reflects the potential impact of a scholarly article; (2) Consideration of scholarly paper impact decaying over time, scholarly papers' quality, early citations, and early citing authors' impact, to quantify the potential impact of scholarly articles; (3) Discussions on how PPI outperforms the existing multi-feature models in citation prediction.
\section{Related work}
Citation prediction of scholarly papers has been extensively investigated, and these studies are mostly based on the analysis of mixture of features, including author-based features (the number of authors, the country of the author's institution, authors' authority, etc.), journal-based features (the total citations of the journal, journal impact factor, keyword frequency of each journal, etc.), paper-based features (the topic of scholarly paper, scholarly paper length, keyword repetition in the abstract of a paper, the number of references, etc.), and other features such as institutional features (institutional rankings and reputation, etc.) In addition, Altmetrics are also employed to predict the citations of scholarly paper. Various investigations have been conducted to explore the correlation between Twitter activities and citation patterns \citep{Peoples2016Twitter,Timilsina2016Towards,Erdt2016Altmetrics}. Seminal examples in citation prediction using mixture of features are summarized in Table \ref{tab:1}. The three categories of features: author-based features, journal-based features, and citations feature are used in our multi-feature predictive model. In order to improve the performance of prediction, Author Impact Factor (AIF), Q value, H-index, Journal Impact Factor and citations are used to predict the citations of scholarly paper. The main difference between our multi-feature predictive model and the prior studies is the selection of features.

%\begin{adjustwidth}{-1cm}{-1cm}
\begin{table*}[!htbp]
  \renewcommand{\arraystretch}{1.2}
\centering
\caption{\bf Examples of multi-feature citation prediction of scholarly paper.}%%%Table caption goes here
\label{tab:1}
\begin{threeparttable}
\begin{tabular}{|m{2.5cm}<{\centering}|m{2.8cm}<{\centering}|m{3cm}<{\centering}|m{2.8cm}<{\centering}| m{2.8cm}<{\centering}| } \hline
 source                    & author features       & journal features   & paper features     & other features \\
    \hline
    \cite{Haslam2008What}     & the number of authors, first author gender & journal prestige & title length, the number of references                       & first author institution's prestige \\ \hline
     \cite{Bornmann2012Factors}& the number of authors, the reputation of the authors     & the language of the publishing journal      & citation count& citation performance of the cited references, reviewers' ratings of importance \\ \hline
    \cite{livne2013predicting}& H-index, g-index              & journal prestige                   & citations     &  content similarity, graph density, clustering coefficient  \\ \hline
     \cite{yu2014citation}       &  the number of authors, the country of the author's institution, H-index                & journal impact factor, total citations, 5-year impact factor, the cited half-life                  & the number of references, the reciprocal of the first-cited age of this paper & the document type\\
    \hline
    \cite{Singh2015The}       &  H-index, author rank, past influence of authors, productivity, sociality, authority, versatility                     & journal rank, journal centrality, past influence of journals                   & publication count, citation count, novelty, topic rank, diversity    & average countX, average citeWords\\
    \hline
  \cite{robson2016can}       &  the number of authors, author name  & the number of journal pages, journal prestige                  & the year of publication, title length, abstract length    & special issue\\
    \hline
 \cite{Sohrabi2017The}       & the number of authors  & & title length, abstract length& SCImago quartile  \\
    \hline
\end{tabular}
\end{threeparttable}
\end{table*}%%%End of the table
%\end{adjustwidth}
In order to analyze the efficiency of multi-feature for citation prediction, regression models are often used. Popular regression models for citation prediction include quantile regression \citep{robson2016can}, semi-continuous regression \citep{Sohrabi2017The} and Gradient Boosted Regression Trees (GBDT) \citep{Chen2015Predicting}. Generative models can also be used to predict the citations of scholarly papers~\citep{Li2015Trend,Zhang2016AdaWIRL}. \cite{wang2013quantifying} proposed a point process by identifying three fundamental mechanisms in paper impact prediction: preferential attachment (highly cited papers are more likely be cited again), decay rate, and fitness (capturing the inherent differences between papers) to predict the probability of a paper being cited. To characterize the citation dynamics of scientific papers, a nonlinear stochastic model of citation dynamics based on the copying-redirection-triadic closure mechanism was reported by \cite{Golosovsky2017Growing}.
\section{Modeling citing behavior as a point process}
\subsection{Dataset}
The American Physical Society (APS) dataset includes all papers published in 9 journals, including Physical Review A, Physical Review B, Physical Review C, Physical Review D, Physical Review E, Physical Review I, Physical Review L, Physical Review ST and Review of Modern Physics, from 1970 to 2013 (http://publish.aps.org/datasets). Each record in the APS dataset includes paper title, author names, author affiliations, date of publication, and a list of cited papers. Because the APS dataset does not provide unique author identifiers, we first do name disambiguation based on the method proposed by \cite{sinatra2016quantifying} in our experiments. Two authors are considered to be the same individual if all of the following three conditions are fulfilled: (1) Last names of two authors are identical; (2) First names are identical or with the matched initial; (3) One of the following is true: the two authors cited each other at least once; the two authors share at least one co-author; The two authors share at least one similar affiliation. We select 183,336 papers as experimental data in the APS dataset from 1978 to 1998. Scholarly papers with greater or equal to 5 citations within the first 5 years of publication are used as the training data, and their citations in the subsequent 10 years are used as the testing data.
\subsection{Prediction model}
\textbf{Intrinsic potential}
Citations reflect the impact of a research paper, which correspond to the authors' impact which can be quantified as $Q_{i}$ for an author $i$~\citep{sinatra2016quantifying}. A scholar with high $Q_{i}$ is expected to publish high-impact publications. In this paper, we use the parameter $Q_{i}$ to indicate the intrinsic potential of a paper's impact.

\begin{figure}[H]
  \centering
  %\centerline{\epsfig{file=networks2.eps,width=\linewidth}}
  \includegraphics[width=0.95\linewidth]{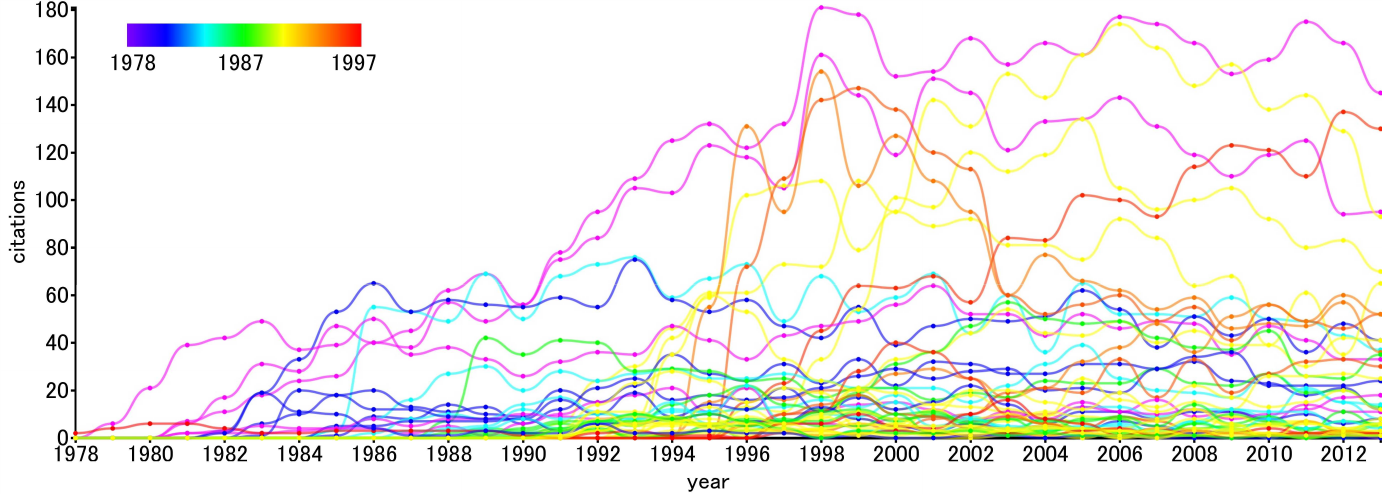}
  \caption{Citation pattern of individual scholarly papers over time.}
  \label{Figure1}
\end{figure}

\textbf{Paper impact decaying over time}
As new ideas presented of each paper further grow in follow-up studies, the novelty fades away eventually and the impact of papers decays over time~\citep{wang2013quantifying}. Figure \ref{Figure1} shows the citation pattern of individual scholarly papers over time. The vertical axis is the yearly citations of 100 randomly selected scholarly papers published between 1978 and 1997 in the APS dataset. The color represents to the publication year of each scholarly paper. According to Figure \ref{Figure1}, each paper has its own inherent citation trend and the pattern may not correlate to one another.

\textbf{Early citers' impact}
Some prior studies have ignored the citers' impact to the citation dynamics \citep{wang2013quantifying}. According to the study in \cite{Singh2017Understanding}, influential early citers might negatively affect long-term scientific impact of papers due to attention stealing, whereas non-influential early citers could positively affect the long-term scientific impact of papers. Inspired by this idea, the early citers' impact is used in PPI to model the citation pattern of a scholarly paper.

\textbf{Early citation}
Based on the behavior that high early citations lead to more citations in the future, we model the Paper Potential Index $\lambda_{d}(t)$ of a scholarly paper $d$ by extending a self-exciting Hawkes process:
\begin{equation}\label{eq:e2}
  \lambda_{d}(t)=\beta_{d} Q_{dMax} e^{-w_{1d}t}+\alpha_{d}\sum_{j,t_{j}<t}D_{j}e^{-w_{2d}(t-t_{j})}
\end{equation}
where parameter $\beta_{d}$ is the coefficient of paper $d$ impact decaying over time. $Q_{dMax}$ indicates the maximum value of authors' impact of paper $d$, and $e^{-w_{1d}t}$ indicates the decay of a paper impact over time. Parameter $\alpha_{d}$ is the coefficient that triggers the current impact of paper $d$. $D_{j}$ indicates the early citers' impact on paper $d$'s citations. $e^{-w_{2d}(t-t_{j})}$ indicates the decay of the current citation.

In equation (\ref{eq:e2}), the $Q$ value reflects an author's influence to the impact of a paper~(\cite{sinatra2016quantifying}), and it is a constant in a scientist's career.
\begin{equation}\label{eq:e3}
Q_{i}=e^{ \left \langle  \log{c_{i\alpha}} \right \rangle - \mu_{p}}
\end{equation}
where $Q_{i}$ represents the $Q$ value of author $i$. $\left \langle  log c_{i\alpha}\right \rangle$ represents the average logarithmic citations of all papers published by author $i$. $\alpha$ represents author $i$'s $\alpha$-th paper. $\mu_{p}$ is equal to $\left \langle \widehat{p} \right \rangle$.

In order to explore the correlation between early citers' impact and paper citations, we conduct experiments based on the method proposed by \citep{Singh2017Understanding}. We first get the maximum $Q$ value among citers for each paper since it is published for two years. Next, we verify the correlation between the maximum $Q$ value for citers with high impact and citations of paper published for 5, 8, 10, 12 and 15 years. We also verify the correlation between the maximum $Q$ value for citers with low impact and citations of paper. Our experimental results show that early citing authors with low impact is more relevant to the long-term scientific impact of papers than early citing authors with high impact. The results are consistent with the finding in \cite{Singh2017Understanding} that attention stealing exists. In accordance with the positive correlation between them, we define $D_{j}$ in equation (2) as:
\begin{equation}\label{eq:e4}
  D_{j}=1+\frac{Q_{j}}{Q_{jMax}}
\end{equation}
where $Q_j$ is the maximum $Q$ value among all authors of a citing paper $j$, and $Q_{jMax}$ represents the highest impact among all citers.
\subsection{Model learning and prediction}
In order to obtain the optimal values of parameters $\alpha$, $\beta$, $w_{1}$, $w_{2}$ in the PPI model, we adopt the maximum likelihood estimation method. Namely, given that the reached probability of the $i-1$th citation at time $t_{i}-1$, we maximize the reached probability of the $i$th citation at time $t_{i}$. The concept can be formulated as follows:
\begin{equation}\label{eq:e5}
  p(t_{i}|t_{i-1})=exp \left( -\int^{t_{i}}_{t_{i-1}}\lambda(t)dt \right) \lambda(t_{i})
\end{equation}
then we use the maximum likelihood estimation method to calculate the likelihood function on the cited sequence of each article, and take the logarithmic function of the maximum likelihood estimate:
\begin{equation}\label{eq:e6}
  log \prod ^{n} _{i=1}p(t_{i}|t_{i-1})=\sum^{n}_{i=1}log\lambda(t_{i})-\int^{T}_{0}\lambda(t)dt
\end{equation}
where $n$ is the citation count of a scholarly paper, $t_{i}$ is the time that the $i-th$ citation occurs, and $T$ is a period of time that a paper is cited. The maximum value of the log-likelihood function is obtained by calculating the minimum of its dual equation. Equation~(\ref{eq:e5}) is brought into the above formula, and add a sparse regularized term $\|\beta\|_{1}$, we get the objective function $L_{\beta}$:
\begin{equation}\label{eq:e7}
  \begin{aligned}
    L_{\beta}=-\sum^{N}_{d=1}\{{\sum^{n}_{i=1}log(\beta s_{d}e^{-w_{1d}t_{i}}+\sum^{i-1}_{j-1}\alpha_{d}D_{j}e^{-w_{2d}(t_{i}-t_{j})}})\\-{\frac{\beta s_{d}}{w_{1d}}(1-e^{-w_{1d}T})-\frac{\alpha_{d}}{w_{2d}}\sum^{n}_{i=1}D_{i}-e^{-w_{2d}(T-t_{i})}}\}+\lambda \|\beta\|_{1}
  \end{aligned}
\end{equation}
where $N$ is the number of papers in the experimental data, $s_{d}$ is the features of a paper. Adding the regularization term makes the objective function non-differentiable, we use the Alternating Direction Method of Multipliers (ADMM) to decompose the original optimization problem into a few simpler sub-problems. By introducing the auxiliary variable $z$, the optimization problem in equation~(\ref{eq:e7}) can be formulated by the following constraint optimization:
\begin{equation}\label{eq:e8}
  \min{L +\lambda\|z\|_{1}} \quad s.t. \quad \beta=z
\end{equation}
The corresponding augmented Lagrangian is:
\begin{equation}\label{eq:e9}
  L_{\rho}=L+\lambda\|z\|_{1}+\rho\mu(\beta-z)+\frac{\rho}{2}\|\beta-z\|^{2}_{2}
\end{equation}
where $\mu$ is the dual variable or Lagrange multiplier; $\rho$ is the penalty coefficient, which is usually used as an iterative step to update the dual variable. The steps to solve the above augmented Lagrange optimization problem using the ADMM algorithm are as follows:
\begin{equation}\label{eq:e10}
  (\beta^{l+1},\alpha^{l+1})=\arg min_{\beta\geq0,\alpha\geq0}{L_{\rho}(\beta^{l},\alpha^{l},z^{l},u^{l})}
\end{equation}
\begin{equation}\label{eq:e11}
  z^{l+1}=S_{\lambda/\rho}(\beta^{l+1}+\alpha^{l+1})
\end{equation}
\begin{equation}\label{eq:e12}
  u^{l+1}=u^{l}+\beta^{l}-z^{l+1}
\end{equation}
where $S_{\lambda/\rho}$ is a soft critical value function. The ADMM algorithm is similar to the dual ascent algorithm, including a parameter minimization process, such as equation~(\ref{eq:e10}); an auxiliary parameter minimization process, such as equation~(\ref{eq:e11}); and a dual parameter update process, such as equation~(\ref{eq:e12}).
In order to efficiently solve the optimization problem in equation~(\ref{eq:e10}), we use the EM framework to update the parameters $\alpha$ and $\beta$. Given the probability that feature $k$ activates event $i$ is $p_{ki}$, the probability that event $i$ activates event $j$ is $p_{ij}$, the EM algorithm is as follows:
\begin{equation}\label{eq:e13}
  p^{d(l+1)}_{ki}=\frac{\beta_{k}s_{dk}e^{-w1dt_{i}}}{\lambda(t_{i})}
\end{equation}
\begin{equation}\label{eq:e14}
  p^{d(l+1)}_{ij}=\frac{\alpha_{d}D_{j}e^{-w_{2d}(t_{i}-t_{j})}}{\lambda(t_{i})}
\end{equation}
\begin{equation}\label{eq:e15}
  \beta^{l+1}_{k}=\frac{-B+\sqrt{B^{2}+4\rho\sum^{N}_{d=1}\sum^{n}_{i=1}p^{d}_{ki}}}{2\rho}
\end{equation}

\begin{equation}\label{eq:e16}
  \alpha^{(l+1)}_{d}=\frac{\sum^{n}_{i=1}\sum^{i-1}_{j=1}p^{d}_{ij}}{\sum^{n}_{i=1}(D_{i}-e^{-w_{2d}(T-t_{i})})/w_{2d}}
\end{equation}
where $B=\sum^{N}_{d=1}s_{dk}(1-e^{-w_{1d}T})/w_{1d}+\rho(u_{k}-z_{k})$. Equation~(\ref{eq:e13}) represents the probability that the value of the $k$th feature $S_{dk}$ and the coefficient $\beta_{k}$ corresponding to the feature $k$ affect the citations of the paper when a paper is cited $i$ times. Equation~(\ref{eq:e14}) represents the probability that the $j$-th ($j \ge i$) citation affects the citations of a paper when it is cited $i$ times. Therefore, $\sum^{N}_{d=1}\sum^{n}_{i=1}\lambda(t_{i})p^{d}_{ki}$ indicates the expectation that the
coefficient $\beta_{k}$ corresponding to the feature $k$ affects citations of the paper on the entire data set. $\sum^{n}_{i=1}\sum^{i-1}_{j=1}\lambda(t_{i})p^{d}_{ij}$ indicates the expectation that the number of existing citations of the paper affects its citations. In equation~(\ref{eq:e9}), we find the maximum of these two expectations and derive the partial derivatives for $\alpha$ and $\beta$. When the partial derivative is zero, equations (\ref{eq:e15}) and~(\ref{eq:e16}) are obtained. By iterating until convergence, we get the optimal values of the parameters $\alpha$ and $\beta$. After that, the new values of $\alpha$ and $\beta$ are brought back to the values of $u$ and $z$ in the ADMM algorithm.

After obtaining the parameters $\alpha$ and $\beta$, the parameters $w_{1}$ and $w_{2}$ of each paper are solved by the gradient descent method. The gradient of the objective function with respect to $w_{1}$ and $w_{2}$ is as follows:
\begin{equation}\label{eq:e17}
  \frac{\partial L_{\rho}}{\partial w_{1}}=\sum^{n}_{i=1}\frac{\beta st_{i}e^{-w_{1}t_{i}}}{\lambda(t_{i})}+\frac{\beta \\ s}{w^{2}_{1}}(e^{-w_{1}T}+T\cdot w_{1}\cdot e^{-w_{1}T}-1)
\end{equation}
\begin{equation}\label{eq:e18}
  \begin{aligned}
    \frac{\partial L_{\rho}}{\partial w_{2}}=&\sum^{n}_{i=1}\frac{\sum^{i-1}_{j=1}(t_{i}-t_{j})\alpha D_{j}e^{-w_{2}(t_{i}-t_{j})}}{\lambda(t_{i})}\\
    &+\frac{\alpha}{w^{2}_{2}}[w_{2}(T-t_{i})e^{-w_{2}(T-t_{i})}+e^{-w_{2}(T-t_{i})}-1]
  \end{aligned}
\end{equation}
After obtaining the optimal values of all parameters $\alpha$, $\beta$, $w_{1}$ and $w_{2}$, we estimate the citations of a scholarly paper after a certain period of time by taking the integral of the intensity function $\lambda(t)$.
\section{Multi-features predictive model}
\subsection{Features that drive the increase of citations}
\noindent
\textbf{\emph{Author-based features}}.
\begin{itemize}
  \item Author Impact Factor (AIF). \\
  Similar to the concept of journal impact factor, an author's AIF in year $T$ is the average citations of published papers in a period of $\Delta T$ years before year $T$. Based on the APS dataset, we compute each author's AIF value according to the author's publishing history and use the statistics of all authors' AIF of a given institution as a group of its features, including sum, maximum, minimum, median, average and deviation. We briefly explore and report the authors' AIF features in this work.
  \item $Q$ value.\\
  The $Q$ value is calculated according to equation \ref{eq:e3}.
  \item H-index.\\
  A scholar has an index value of $H$ if the scholar has $H$ papers with at least $H$ citations. H-index can give an estimate of the impact of a scholar's cumulative research contributions.
\end{itemize}

\noindent
\textbf{\emph{Journal-based feature}}.\\
Journal Impact Factor is a quantitative index to evaluate the impact of journal. It is actually the ratio of citations of a journal and papers published of the journal.\\

\noindent
\textbf{\emph{Citations feature}}.\\
The historical citations of each paper are used to predict the impact of a paper.
\subsection{Feature selection}
In order to investigate the effect of author-based feature, journal-based feature and citations feature, we evaluate the importance of features (see Table \ref{tab:2}).
\begin{table*}[htbp]
  \renewcommand{\arraystretch}{1.2}
  \centering
  \caption{\bf Features used in the prediction model.}
  \begin{threeparttable}
    \begin{tabular}{|c|c|c|c|c|c|c|c|}  \hline
      Feature     & Description         & Feature     & Description\\ \hline
      c1          & one-year citations  &max(H-index) & maximum of H-index  \\ \hline
      c2          & two-year citations  & min(H-index)& minimum of H-index  \\ \hline
      c3          & three-year citations& avg(H-index)& average of H-index \\ \hline
      c4          & four-year citations & med(H-index)& median of H-index \\ \hline
      c5          & five-year citations & dev(H-index)& deviation of H-index \\ \hline
      sum(Q)      & sum of Q value      & sum(AIF)    & sum of AIF \\ \hline
      max(Q)      & maximum of Q value  & max(AIF)    & maximum of AIF \\ \hline
      min(Q)      & minimum of Q value  & min(AIF)    & minimum of AIF \\ \hline
      avg(Q)      & average of Q value  & avg(AIF)    & average of AIF\\ \hline
      med(Q)      & median of  Q value  & med(AIF)    & median of AIF\\ \hline
      dev(Q)      & deviation of Q value& dev(AIF)    & deviation of AIF \\ \hline
      sum(H-index)& sum of H-index      &JIF          & journal impact factor\\ \hline
    \end{tabular}
  \end{threeparttable}
  \label{tab:2}
\end{table*}
\subsection{Learning algorithm}
In this section, we describe the multi-feature predictive model, which integrates author-based feature, journal-based feature and citations to the Gradient boosting decision trees (GBDT). The GBDT model suits for a mass of features and no-linear relationships between the predictor variables and the target variable. In terms of the multi-feature predictive model, parameters adjustment is crucial for the performance of predictive model. Main parameters include:\\
(1) \textit{learning rate}: namely the model's learning speed on the distribution characteristics of the sample, expressed as the weight of the regression tree for each iteration in the algorithm. The larger the learning rate is, the faster the algorithm converges. The smaller the learning rate is, the slower the algorithm converges, but the prediction accuracy may increase.\\
(2) \textit{number of iterations}: the number of iterations is the number of weak learners obtained in the model. In general, the number of iterations depends on the learning rate.\\
(3) \textit{minimum samples of leaf nodes}: this parameter defines the conditions under which the subtree continues to be divided. If the number of samples on the leaf node is smaller than the set value, the node will not be further divided.\\
(4) \textit{maximum depth of decision tree}: this parameter is used to control the maximum depth of the decision tree generated by each round iteration. The purpose is to prevent over-fitting.\\
(5) \textit{Sampling rate}: this parameter indicates the proportion of training samples used in each training, and its value ranges from 0 to 1. When the value is 1, it indicates that all the samples are involved training. The main role of this parameter is to add sample perturbation to prevent over-fitting. The sampling rate of general samples is set between 0.5 and 0.8. If the value is too large, the risk of over-fitting will be increased. If the value is too small, correct samples may not be learned due to too few samples, and the model deviations will increase.

We used the Grid Search method to adjust the above mentioned parameters. The value of the learning rate ranges from 0.0005 to 0.5 and the step size is 0.0005. The number of iterations ranges from 500 to 3000 and the step size is 500. The value of leaf node minimum sample number value ranges from 10 to 80 and the step size is 10. The maximum depth of the decision tree  ranges from 5 to 7 and the step size is 1. Sampling rate ranges from 0.5 to 1.0 and the step size is 0.1.

According to the range of values and the step size of each parameter, the grid covered parameter space is generated for grid search. Each point on the grid is traversed, and the parameter combination corresponding to the point is used to train the model on the training set. Correspondingly, prediction is performed on the validation set, and the predictive accuracy is calculated as an estimate of the prediction performance of the model under the set of parameters. After traversing all the parameter combinations, the set of parameters with the highest prediction accuracy on the corresponding verification set is taken as the parameter of the final model.
\section{Results and discussion}
\subsection{Evaluation metrics}
In this subsection, we introduce several evaluation metrics for validating the PPI prediction model.\\
\textbf{\emph {Mean absolute error (MAE)}}.\\
MAE quantifies how close the predictions is to the ground truth. MAE is given by:
\begin{equation}
 MAE=\frac{1}{n}\sum_{i=1}^{n}\left |{e_{i}}  \right |
\end{equation}
The mean absolute error is an average of the absolute errors $\left |{e_{i}}  \right |$ , which is equal to $\left |f_{i} -y_{i}\right |$, where $f_{i}$ is the prediction, and $y_{i}$ is the true value. $n$ represents the number of predictions.\\
\textbf{\emph {Root mean squared error (RMSE)}}.\\
RMSE is similar to MAE, which is defined as follows:
\begin{equation}
 RMSE=\sqrt{\frac{1}{n}\sum_{i=1}^{n}{e_{i}}^{2}}
\end{equation}
RMSE also provides the average error and quantify the overall error rate. In some cases, we need to compare results across activities, but RMSE can not give an indication of the relative error. We need a normalized error, such as Range-normalized RMSE.\\
\textbf{\emph{Range-normalized RMSE (NRMSE)}}.\\
\begin{equation}
  NRMSE=\frac{RMSE}{max(y_{i})-min(y_{i})}
\end{equation}
where $max(y_{i})$ and $min(y_{i})$ represent the maximum and minimum functions, which are calculated by all ground-truth values of the test instances.\\
\textbf{\emph {Mean absolute percentage error (MAPE)}}\\
An useful normalized metric is MAPE, which normalizes each error value for each prediction. This metric shows the average deviation between predicted output and true output from the $n$ experimental data. MAPE is defined as follows:
\begin{equation}
  MAPE=\frac{1}{n} \sum_{i=1}^n \frac{\left |e_{i}  \right |}{y_{i}}
\end{equation}
\textbf{\emph{Accuracy}}\\
Accuracy shows the fraction of papers correctly predicted for a given error tolerance $\epsilon$:
\begin{equation}
  Accuracy=\frac{1}{n} \sum_{i=1}^n \mid\frac{\left |e_{i}  \right |}{y_{i}}\leq\epsilon\mid
\end{equation}
\subsection{Feature importance analysis}
\begin{figure}[H]
  \centering
  \includegraphics[width=0.60\linewidth]{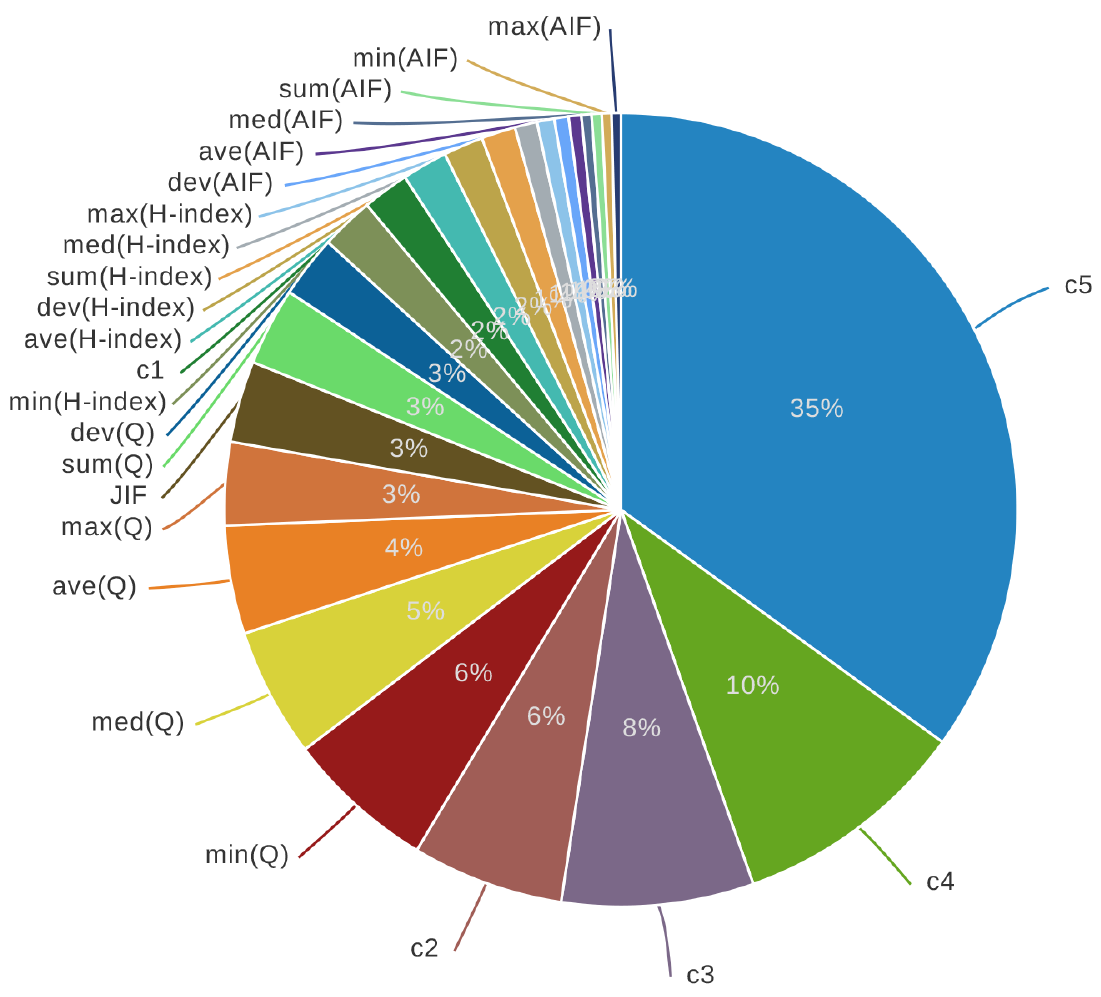}
  \caption{Feature importance score of all features.}
  \label{Figure2}
\end{figure}

Figure \ref{Figure2} shows the feature importance score of all features to predict the 15th year's citations of the published papers. The features $c5$, $c4$, $c3$ and $c2$ rank first to fourth in the feature importance rankings, and their values are 0.3495, 0.0963, 0.0783 and 0.0618. The minimum, median, average, maximum of authors' $Q$ value, JIF, authors' $Q$ value' sum, respectively, their values are 0.0608, 0.0527, 0.0441, 0.0338, 0.0331, and 0.0317. The feature importance score for predicting 6th-14th year citations of the published papers are shown in the appendix.

Based on all features' importance in predicting 6th-15th year citations of papers, we selected the top 10 feature retraining model, the prediction accuracy remains high. There are differences in feature importance scores for different predictive years (see appendix). Figure \ref{Figure3} shows the top 10 feature importance score for predicting the 15th year citations of the published papers. The features $c5$ and $c4$ rank first to third in the feature importance rankings. Their importance scores are 0.3425 and 0.1079, respectively. The authors' $Q$ value's minimum ranks fourth, and its value is 0.0969. Other feature importance scores are less than 0.0950.

\begin{figure}[H]
  \centering
  \includegraphics[width=0.60\linewidth]{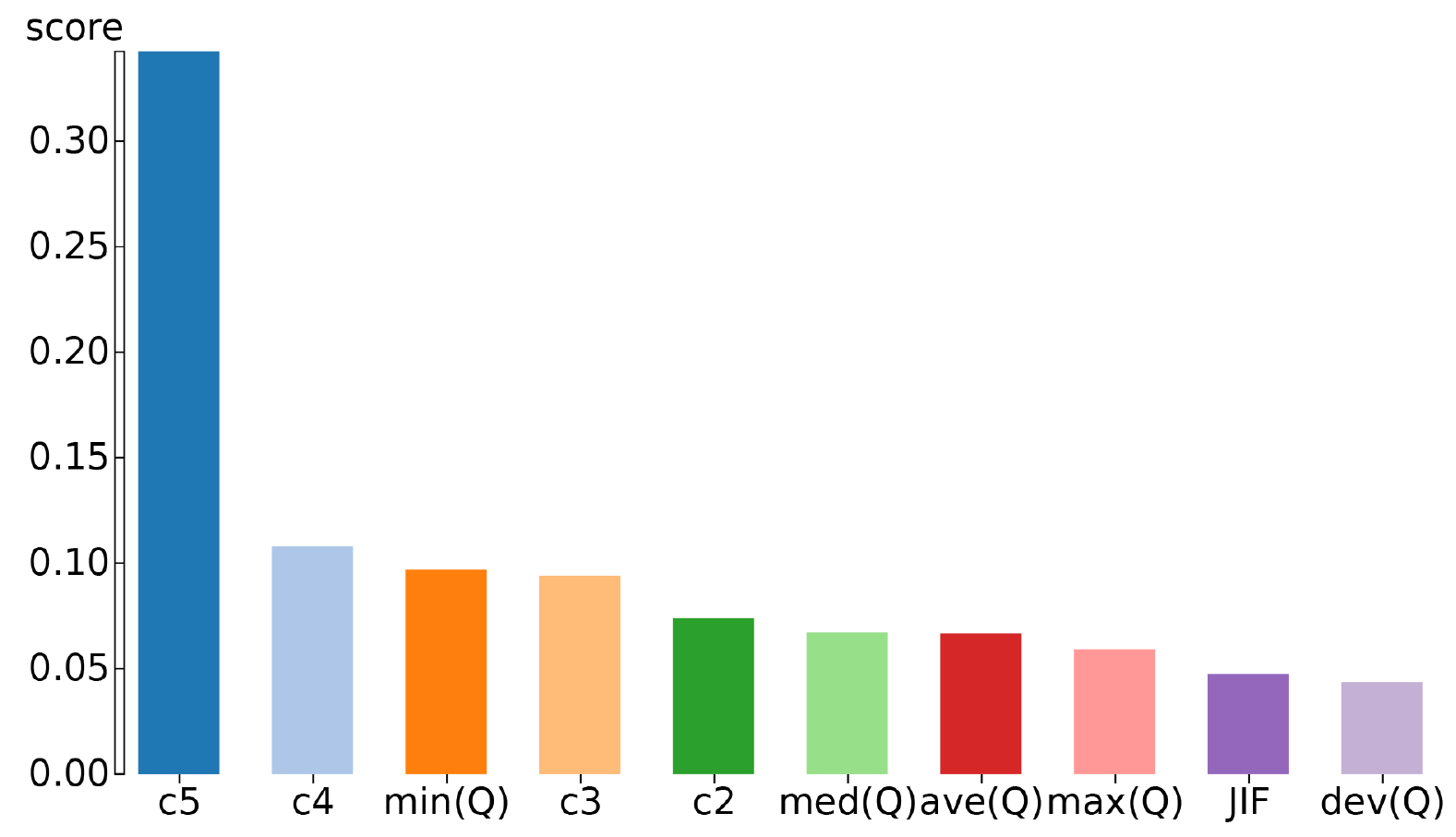}
  \caption{Importance scores of top 10 features.}
  \label{Figure3}
\end{figure}

According to Figure \ref{Figure2} and Figure \ref{Figure3}, we observe that citations in the first five years after the publication of the paper, author's $Q$ value, and JIF are ranked in the top 10 of the list of feature importance ranking. Author H-index related features and author AIF related features are located behind the list of feature importance rankings.

In summary, we observe that historical citations play an important role for predicting the impact of the paper. Besides, author-based features are important in predicting the paper impact, especially the authors' $Q$ value.
\subsection{Comparing performances of different models and discussion}
To test the validity of PPI prediction model, its predictive performance is compared against four competing models: PPI\_NECAI, GBDT\_All, GBDT\_10 and PLI\_Science published by \cite{wang2013quantifying}. The comparison is made in terms of MAE, RMSE, NRMSE, MAPE, and accuracy.

\begin{figure}[!htp]
  \centering
  \includegraphics[width=0.70\linewidth]{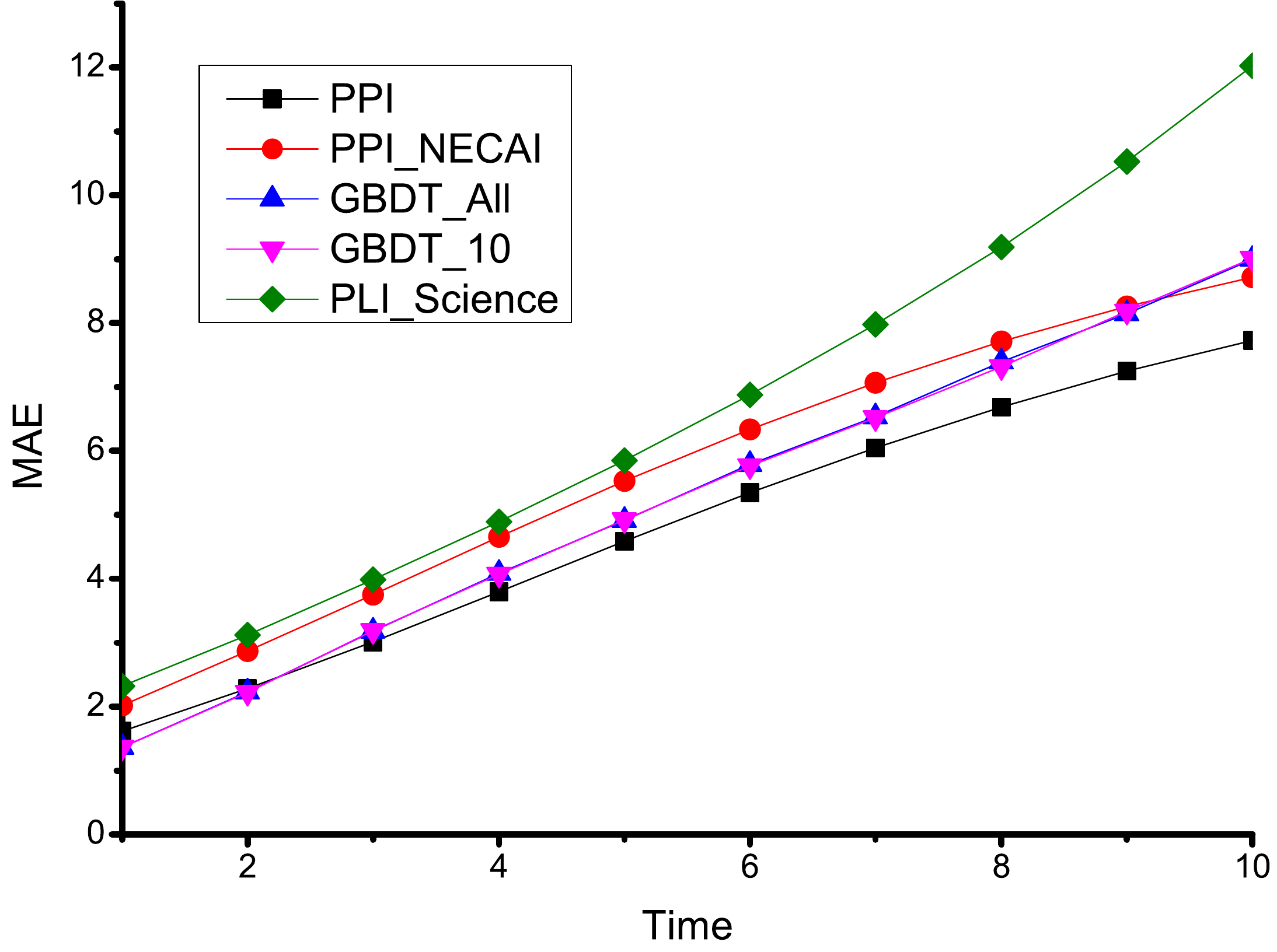}
  \caption{Comparing MAE for different models.}
  \label{Fig. 4}
\end{figure}

Figure \ref{Fig. 4} shows the MAE value of the five models. According to Figure \ref{Fig. 4}, we observe that PPI outperforms all competing models with lower MAE values for predicting citations after a scholarly paper is published for 5 years. We also observe that MAE values of all five models increase along with the year, indicating that the predictive performance of all five models degrades over time.

\begin{figure}[!htp]
  \centering
  \includegraphics[width=0.70\linewidth]{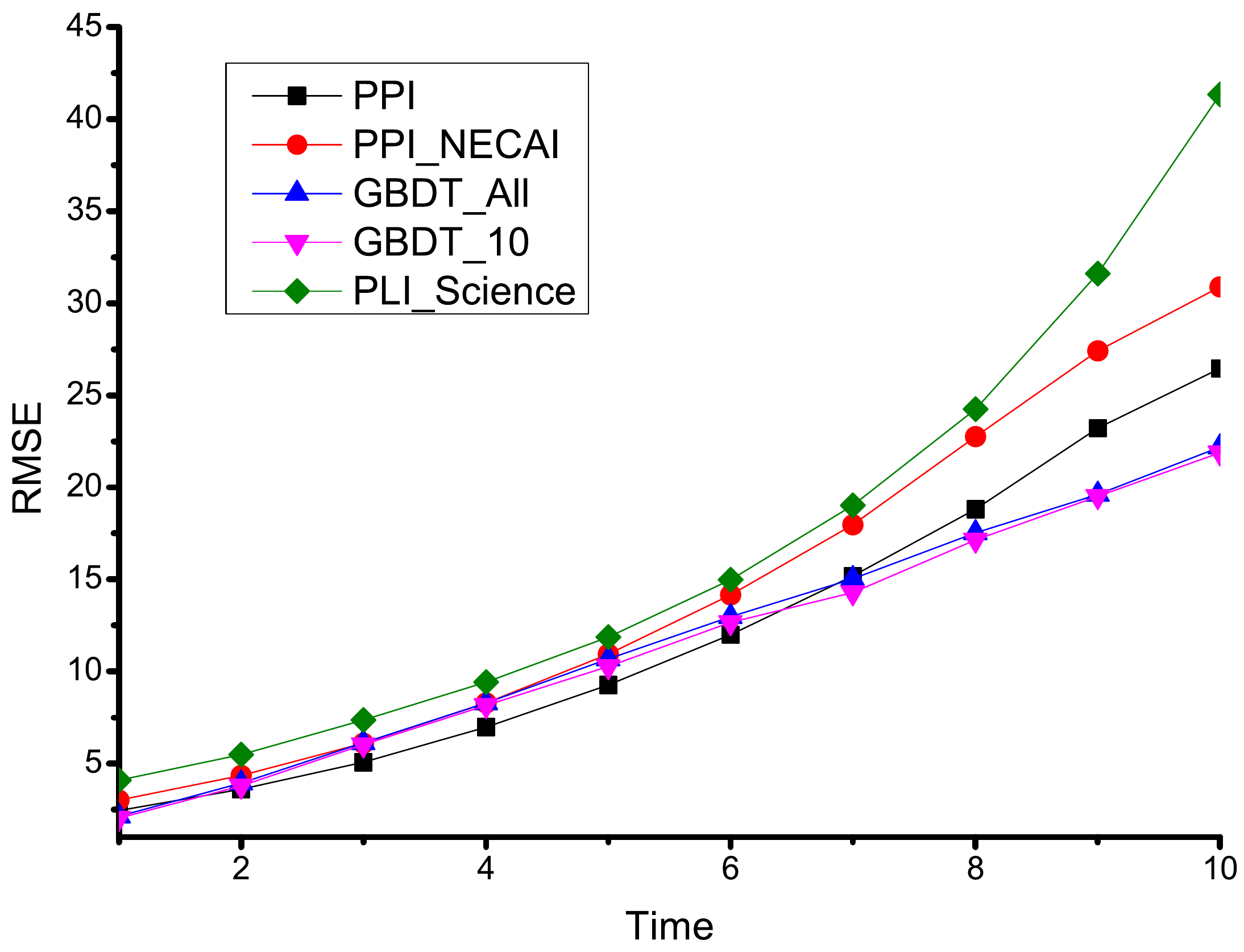}
  \caption{Comparing RMSE for different models.}
  \label{Fig. 5}
\end{figure}

Figure \ref{Fig. 5} shows the RMSE value of the five models. Similar to the MAE comparisons, that the PLI\_Science model falls behind all competing models in terms of RMSE. RMSE performance of all other models are mixed, with PPI yields lower RMSE than other models between 2 to 6 years, indicating it performs well in short term citation prediction but its performance fails behind GBDT\_All and GBDT\_10 for long term citation prediction. Similar to study based on MAE, the predictive performance in terms of RMSE of all five models gradually declines over time.

\begin{figure}[!htp]
  \centering
  %\centerline{\epsfig{file=networks2.eps,width=\linewidth}}
  \includegraphics[width=0.70\linewidth]{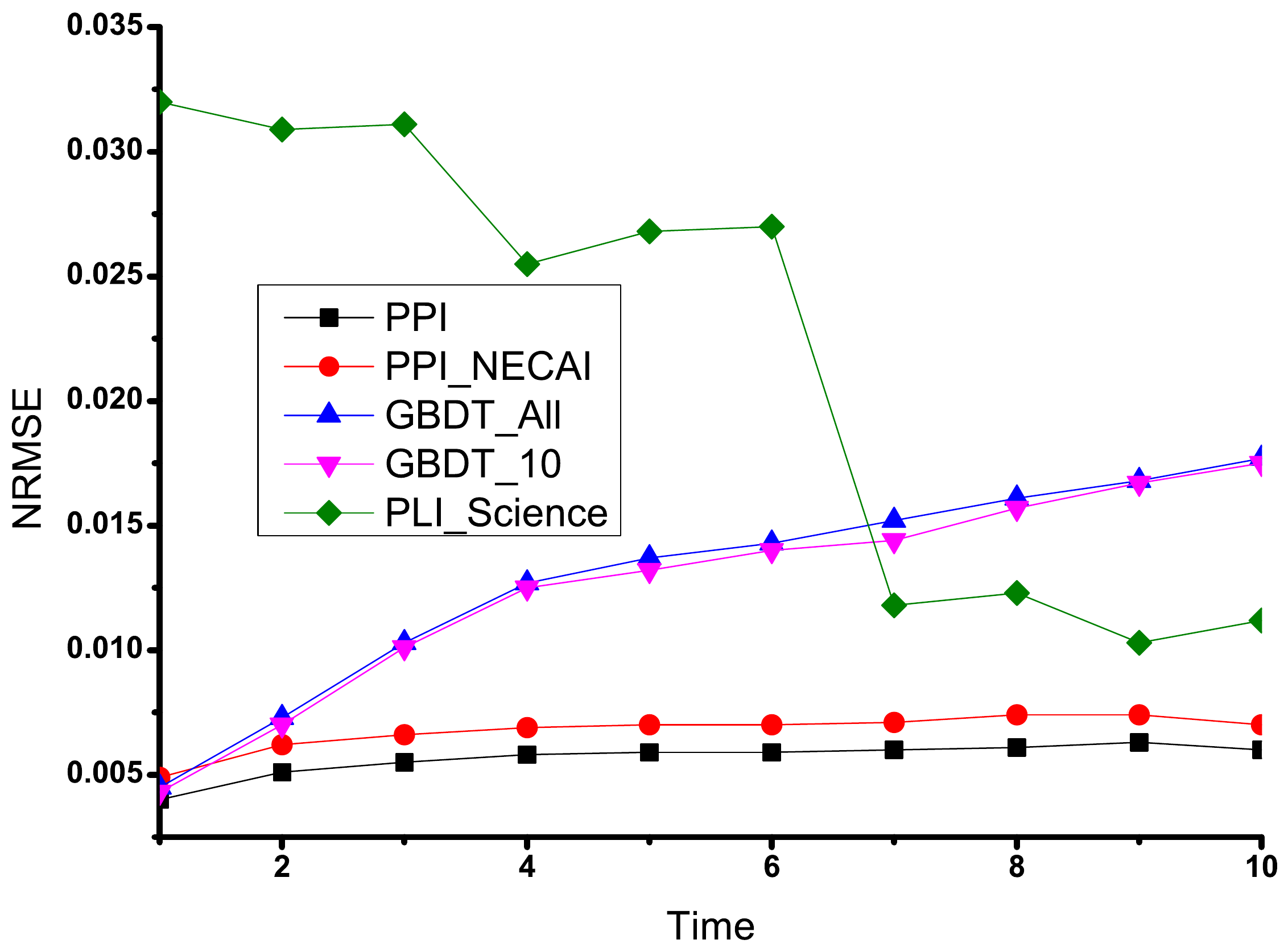}
  \caption{Comparing NRMSE for different models.}
  \label{Fig. 6}
\end{figure}

Figure \ref{Fig. 6} shows NRMSE values of the five models. For PPI model and PPI\_NECAI model, their NRMSE values are about 0.006. The NRMSE values of GBDT\_All model and GBDT\_10 model shows increasing trends, and their NRMSE values are about 0.018 in future the 10th years after the fifth year of scholarly paper published. The NRMSE values of the PLI\_Science model show a decaying trend. In term of NRMSE, the predictive performance of the PPI model is better than other four models.

\begin{figure}[!htp]
  \centering
  %\centerline{\epsfig{file=networks2.eps,width=\linewidth}}
  \includegraphics[width=0.70\linewidth]{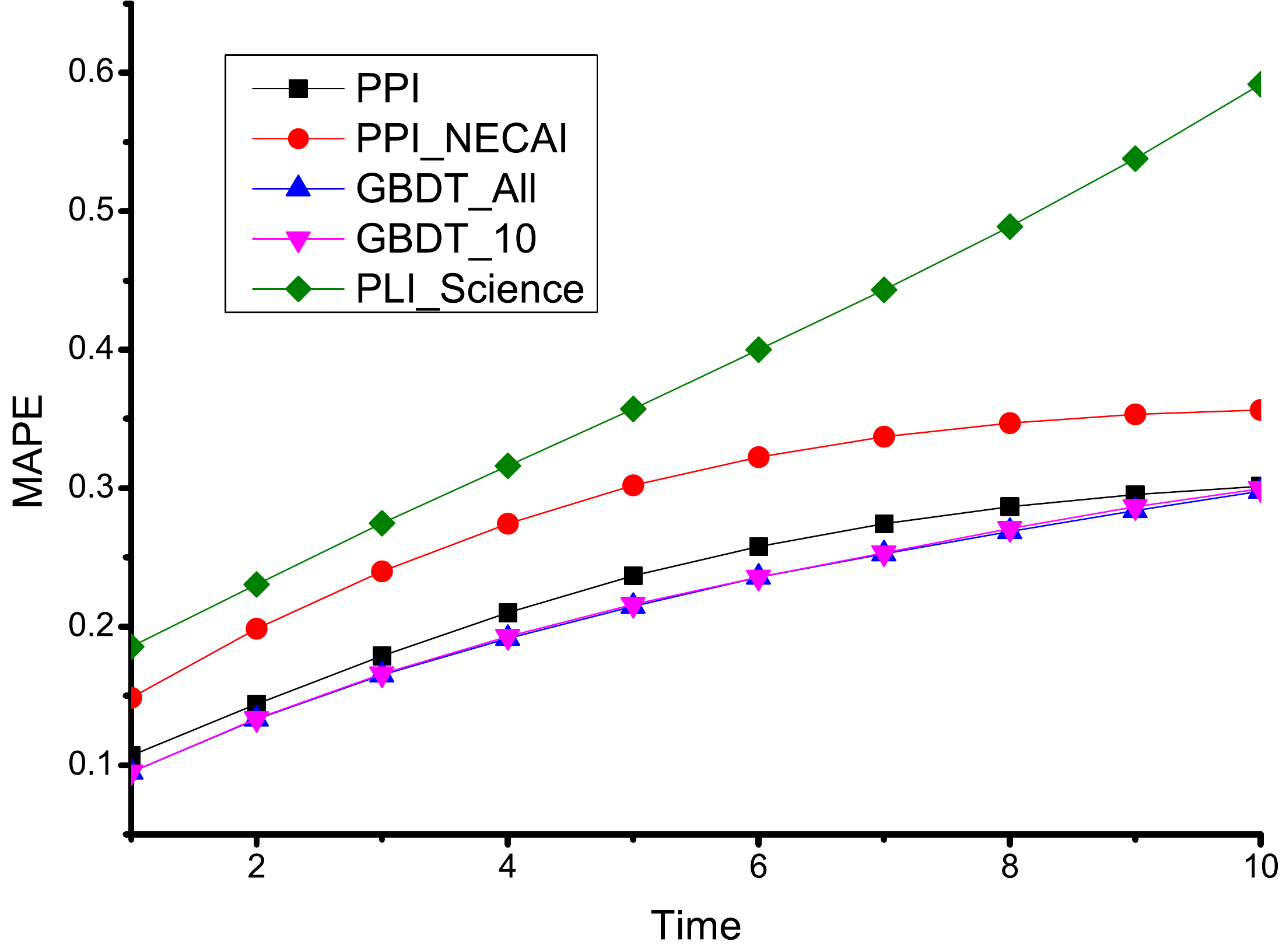}
  \caption{Comparing  MAPE for different models.}
  \label{Fig. 7}
\end{figure}

Figure \ref{Fig. 7} shows the MAPE values of the five models. We observe that the MAPE values of GBDT\_All model and GBDT\_10 model are below the other three models. The MAPE value of the PPI model is slightly higher than GBDT\_All model and GBDT\_10 model.

\begin{figure}[!htp]
  \centering
  %\centerline{\epsfig{file=networks2.eps,width=\linewidth}}
  \includegraphics[width=0.70\linewidth]{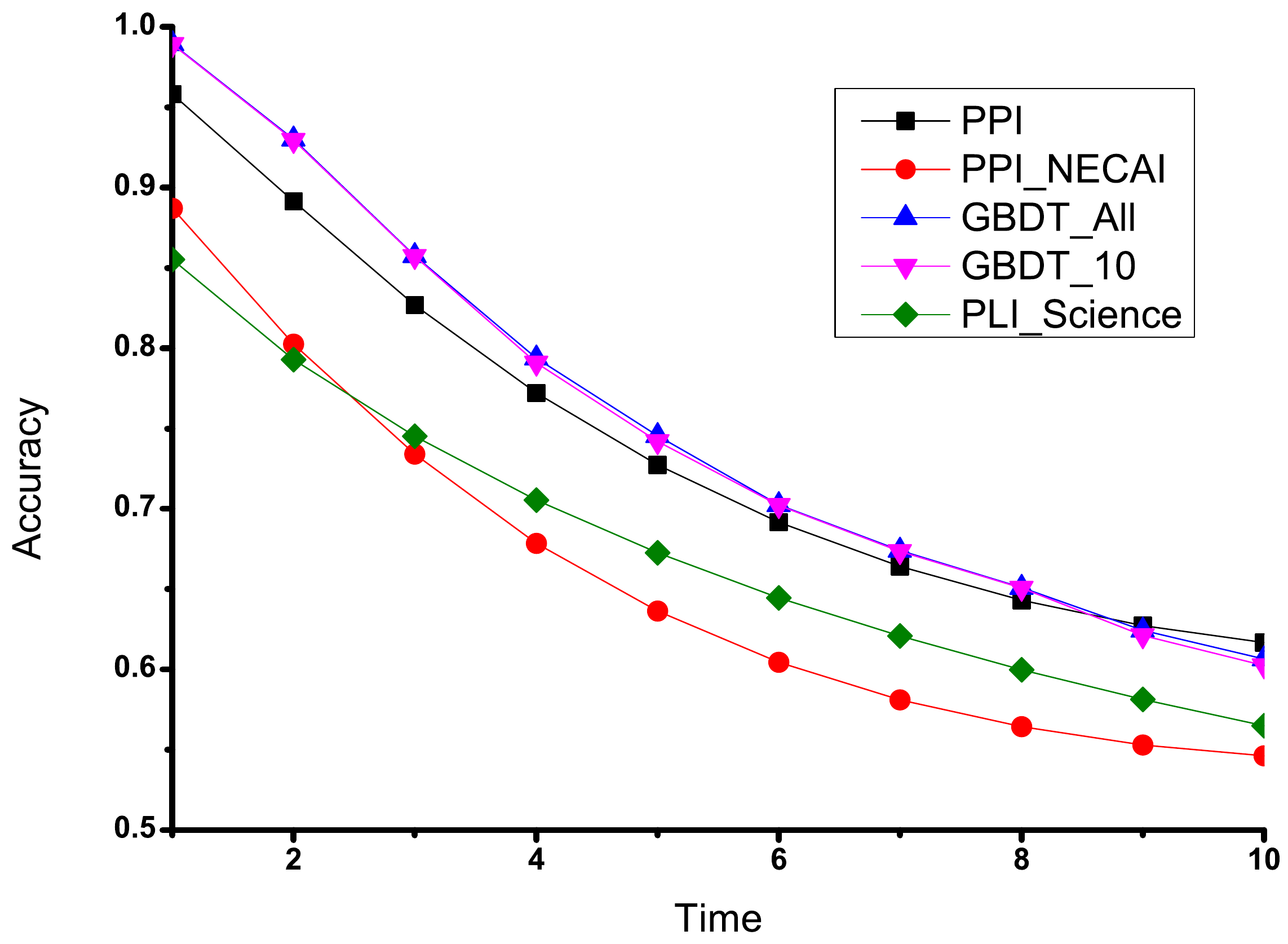}
  \caption{Comparing Accuracy for different models.}
  \label{Fig. 8}
\end{figure}

Figure \ref{Fig. 8} shows the accuracy of the five models. The accuracy values of PPI prediction model are higher than PPI\_NECAI model and PLI\_Science model, but are slightly below than GBDT\_All model and GBDT\_10 model. From 3 to 10 years after the fifth year of scholarly paper published, the predictive accuracy of PPI\_NECAI model is lower than other four models. The predictive accuracy of all models shows a decaying trend.

By comparing PPI and PPI-NECAI, we observe that early citing authors' impact contributes to improved prediction of scholarly paper impact. PPI yields superior citation prediction over PPI-NECAI, GBDT\_All, GBDT\_10 and PLI-Science in terms of MAE and NRMSE.  Although the predictive performances of the GBDT\_All model and the GBDT\_10 model are better than other three models in terms of MAPE and accuracy, the proposed PPI prediction model gives a clear explanation for the predictive effect of the model by the following factors: inherent quality of scholarly paper, scholarly paper impact decaying over time, early citations, and early citers' impact.

Compared to PPI-NECAI and PLI\_Science, PPI more accurately predicts the scholarly paper impact. Although considering early citers' impact can improve the predictive performance of PPI model, other factors exist, such as author's team impact, journal impact, authors' cooperation relationship, and disciplinary differences. In addition, due to the fact that the APS dataset only contains local citations, this might limit the predictive accuracy of this work. Uncovering the essence of paper potential index is a promising future work, which might improve the predictive performance of PPI model, and it could provide a better understanding of the evolution of scholarly paper impact.
\section{Conclusion}
Based on point estimation process, we present the PPI predictive model, which considers the following four factors: (1) inherent quality of scholarly paper; (2) scholarly paper impact decaying over time; (3) early citations; and (4) early citers' impact. Experimental results indicate that the PPI model improves citation prediction of scholarly papers. The predictive performance of PPI is better than PPI-NECAI, which reflects that early citing author's impact is important for predicting the citations of scholarly paper. Although the predictive performance of the GBDT\_All model and GBDT\_10 model is better than other three models in terms of MAPE and accuracy, the proposed PPI predictive model give a clear explanation for the predictive effect, indicating that an ultimate understanding of long-term impact of scholarly paper will benefit from understanding the inherent evolutionary mechanism of citations of scholarly papers.

\section*{Acknowledgement}
We thank Feng Xia and Jie Hou from School of Software, Dalian University of Technology for valuable discussions on this work. This work was partially supported by Liaoning Provincial Key R\&D Guidance Project (2018104021) and Liaoning Provincial Natural Fund Guidance Plan (20180550011).

% \section*{References}
%\bibliography{sample}

\begin{thebibliography}{00}
\expandafter\ifx\csname natexlab\endcsname\relax\def\natexlab#1{#1}\fi
\expandafter\ifx\csname url\endcsname\relax
  \def\url#1{\texttt{#1}}\fi
\expandafter\ifx\csname urlprefix\endcsname\relax\def\urlprefix{URL }\fi

\bibitem[{Adankon and Cheriet(2010)}]{Adankon2010Support}
Adankon, M.~M., Cheriet, M. (2010). Support vector machine. In: International
  Conference on Intelligent Networks and Intelligent Systems. pp. 418--421.

\bibitem[{Bai et~al.(2017)Bai, Liu, Zhang, Ning, Kong, Lee, and
  Xia}]{Bai2017An}
Bai, X., Liu, H., Zhang, F., Ning, Z., Kong, X., Lee, I., Xia, F. (2017). An
  overview on evaluating and predicting scholarly article impact. Information
  8(3), 73.

\bibitem[{Bai et~al.(2016)Bai, Xia, Lee, Zhang, and Ning}]{Bai2016Identifying}
Bai, X., Xia, F., Lee, I., Zhang, J., Ning, Z. (2016). Identifying anomalous
  citations for objective evaluation of scholarly article impact. Plos One
  11(9), e0162364.

\bibitem[{bornmann et~al.(2012)Bornmann, Schier, Marx, and Daniel}]{Bornmann2012Factors}
Bornmann, L., Schier, H., Marx, W., Daniel, H. (2012). What factors determine citation counts of publications in chemistry besides their quality? Journal of Informetrics
  6(1), 11--18.

\bibitem[{Cao et~al.(2016)Cao, Chen, and Liu}]{Cao2016A}
Cao, X., Chen, Y., Liu, K. J.~R. (2016). A data analytic approach to quantifying
  scientific impact. Journal of Informetrics 10(2), 471--484.

\bibitem[{Chen and Zhang(2015)}]{Chen2015Predicting}
Chen, J., Zhang, C. (2015). Predicting citation counts of papers. In: IEEE
  International Conference on Cognitive Informatics \& Cognitive Computing. pp.
  434--440.

\bibitem[{Chen and Guestrin(2016)}]{Chen2016XGBoost}
Chen, T., Guestrin, C. (2016). Xgboost: A scalable tree boosting system. In: ACM
  SIGKDD International Conference on Knowledge Discovery and Data Mining. pp. 785--794.

\bibitem[{Clauset et~al.(2017)Clauset, Larremore, and
  Sinatra}]{clauset2017data}
Clauset, A., Larremore, D.~B., Sinatra, R. (2017). Data-driven predictions in
  the science of science. Science 355(6324), 477--480.

\bibitem[{Erdt et~al.(2016)Erdt, Nagarajan, Sin, and
  Theng}]{Erdt2016Altmetrics}
Erdt, M., Nagarajan, A., Sin, S. C.~J., Theng, Y.~L. (2016). Altmetrics: an
  analysis of the state-of-the-art in measuring research impact on social
  media. Scientometrics 109(2), 1117--1166.

\bibitem[{Fiala and Tutoky(2018)}]{Fiala2018PageRank}
Fiala, D., Tutoky, G. (2018). Pagerank-based prediction of award-winning
  researchers and the impact of citations. Journal of Informetrics 11(4),
  1044--1068.

\bibitem[{Golosovsky and Solomon(2017)}]{Golosovsky2017Growing}
Golosovsky, M., Solomon, S. (2017). Growing complex network of citations of
  scientific papers: Modeling and measurements. Phys.rev.e 95(1): 012324.

\bibitem[{Haslam et~al.(2008)Haslam, Ban, Kaufmann, Loughnan, Peters, Whelan,
  and Wilson}]{Haslam2008What}
Haslam, N., Ban, L., Kaufmann, L., Loughnan, S., Peters, K., Whelan, J.,
  Wilson, S. (2008). What makes an article influential? predicting impact in
  social and personality psychology. Scientometrics 76(1), 169--185.

\bibitem[{Hawkes and Oakes(1974)}]{Hawkes1974A}
Hawkes, A.~G., Oakes, D. (1974). A cluster process representation of a
  self-exciting process. Journal of Applied Probability 11(3), 493--503.

\bibitem[{Heidi(2015)}]{Ledford2015Team}
Heidi, L. (2015). Team science. Nature 525(7569), 308--222.

\bibitem[{Li et~al.(2015)Li, Lin, Yan, and Yeh}]{Li2015Trend}
Li, C.~T., Lin, Y.~J., Yan, R., Yeh, M.~Y. (2015). Trend-Based Citation Count
  Prediction for Research Articles. In: Springer, Cham, pp. 659-671.

\bibitem[{Liu et~al.(2017)Liu, Yan, Xiao, Wang, Zha, and Chu}]{liu2017on}
Liu, X., Yan, J., Xiao, S., Wang, X., Zha, h., Chu, S.~M. (2017). On predictive
  patent valuation: Forecasting patent citations and their types. In: AAAI. pp.
  1438--1444.

\bibitem[{Livne et~al.(2013)Livne, Adar, Teevan, and
  Dumais}]{livne2013predicting}
Livne, A., Adar, E., Teevan, J., Dumais, S. (2013). Predicting citation counts
  using text and graph mining. In: Proc. the iConference 2013 Workshop on
  Computational Scientometrics: Theory and Applications. pp. 1-4.

\bibitem[{Newman(2008)}]{Newman2008The}
Newman, M. E.~J. (2008). The first-mover advantage in scientific publication.
  Epl 86(6), 68001--68006.

\bibitem[{Panagopoulos et~al.(2017)Panagopoulos, Tsatsaronis, and
  Varlamis}]{Panagopoulos2017Detecting}
Panagopoulos, G., Tsatsaronis, G., Varlamis, I. (2017). Detecting rising stars
  in dynamic collaborative networks. Journal of Informetrics 11(1), 198--222.

\bibitem[{Peoples et~al.(2016)Peoples, Midway, Sackett, Lynch, and
  Cooney}]{Peoples2016Twitter}
Peoples, B.~K., Midway, S.~R., Sackett, D., Lynch, A., Cooney, P.~B. (2016).
 twitter predicts citation rates of ecological research. Plos One 11(11),
  e0166570.

\bibitem[{Petersen et~al.(2014)Petersen, Fortunato, Pan, Kaski, Penner, Rungi,
  Riccaboni, Stanley, and Pammolli}]{petersen2014reputation}
Petersen, A.~M., Fortunato, S., Pan, R.~K., Kaski, K., Penner, O., Rungi, A.,
  Riccaboni, M., Stanley, H.~E., Pammolli, F. (2014). Reputation and impact in
  academic careers. Proceedings of the National Academy of Sciences 111(43),
  15316--15321.

\bibitem[{Pobiedina and Ichise(2016)}]{Pobiedina2016Citation}
Pobiedina, N., Ichise, R. (2016). Citation count prediction as a link prediction
  problem. Applied Intelligence 44(2), 252--268.

\bibitem[{Robson and Mousqu{\`e}s(2016)}]{robson2016can}
Robson, B.~J., Mousqu{\`e}s, A. (2016). Can we predict citation counts of
  environmental modelling papers? Fourteen bibliographic and categorical
  variables predict less than 30\% of the variability in citation counts.
  Environmental Modelling \& Software 75, 94--104.

\bibitem[{Sandulescu and Chiru(2016)}]{sandulescu2016predicting}
Sandulescu, V., Chiru, M. (2016). Predicting the future relevance of research
  institutions-the winning solution of the KDD Cup 2016. arXiv preprint
  arXiv:1609.02728.

\bibitem[{Sarig{\"o}l et~al.(2014)Sarig{\"o}l, Pfitzner, Scholtes, Garas, and
  Schweitzer}]{Sarig2014Predicting}
Sarig{\"o}l, E., Pfitzner, R., Scholtes, I., Garas, A., Schweitzer, F. (2014).
  Predicting scientific success based on coauthorship networks. EPJ Data
  Science 3(1), 9.

\bibitem[{Sinatra et~al.(2016)Sinatra, Wang, Deville, Song, and
  Barab{\'a}si}]{sinatra2016quantifying}
Sinatra, R., Wang, D., Deville, P., Song, C., Barab{\'a}si, A.-L. (2016).
  Quantifying the evolution of individual scientific impact. Science
  354(6312), aaf5239.

\bibitem[{Singh et~al.(2017)Singh, Jaiswal, Shree, Pal, Mukherjee, and
  Goyal}]{Singh2017Understanding}
Singh, M., Jaiswal, A., Shree, P., Pal, A., Mukherjee, A., Goyal, P. (2017).
  Understanding the impact of early citers on long-term scientific impact. In:
  Digital Libraries. pp.59--68.

\bibitem[{Singh et~al.(2015)Singh, Patidar, Kumar, Chakraborty, Mukherjee, and
  Goyal}]{Singh2015The}
Singh, M., Patidar, V., Kumar, S., Chakraborty, T., Mukherjee, A., Goyal, P.,
  (2015). The role of citation context in predicting long-term citation profiles:
  An experimental study based on a massive bibliographic text dataset. In: ACM
  Conference on Information \& Knowledge Management. pp. 1271--1280.

\bibitem[{Sohrabi and Iraj(2017)}]{Sohrabi2017The}
Sohrabi, B., Iraj, H. (2017). The effect of keyword repetition in abstract and
  keyword frequency per journal in predicting citation counts. Scientometrics
  110(1), 1--9.

\bibitem[{Stegehuis et~al.(2015)Stegehuis, Litvak, and
  Waltman}]{stegehuis2015predicting}
Stegehuis, C., Litvak, N., Waltman, L. (2015). Predicting the long-term citation
  impact of recent publications. Journal of informetrics 9(3), 642--657.

\bibitem[{Tahamtan et~al.(2016)Tahamtan, Safipour~Afshar, and
  Ahamdzadeh}]{Tahamtan2016Factors}
Tahamtan, I., Safipour~Afshar, A., Ahamdzadeh, K. (2016). Factors affecting
  number of citations: a comprehensive review of the literature. Scientometrics
  107(3), 1195--1225.

\bibitem[{Timilsina et~al.(2016)Timilsina, Davis, Taylor, and
  Hayes}]{Timilsina2016Towards}
Timilsina, M., Davis, B., Taylor, M., Hayes, C. (2016). Towards predicting
  academic impact from mainstream news and weblogs: A heterogeneous graph based
  approach. In: IEEE/ACM International Conference on Advances in Social
  Networks Analysis and Mining. pp.1388--1389.

\bibitem[{Wang et~al.(2013)Wang, Song, and Barab{\'a}si}]{wang2013quantifying}
Wang, D., Song, C., Barab{\'a}si, A.-L. (2013). Quantifying long-term scientific
  impact. Science 342(6154), 127--132.

\bibitem[{Wang et~al.(2008)Wang, Yu, and Yu}]{Wang2008Measuring}
Wang, M., Yu, G., Yu, D. (2008). Measuring the preferential attachment mechanism
  in citation networks. Physica A Statistical Mechanics \& Its Applications
  387(18), 4692--4698.

\bibitem[{Xia et~al.(2016)Xia, Su, Wang, Zhang, Ning, and
  Lee}]{Feng2016Bibliographic}
Xia, F., Su, X., Wang, W., Zhang, C., Ning, Z., Lee, I. (2016). Bibliographic
  analysis ofnaturebased on Twitter and facebook altmetrics data:. Plos One
  11(12), e0165997.

\bibitem[{Xia et~al.(2017)Xia, Wang, Bekele, and Liu}]{Xia2017Big}
Xia, F., Wang, W., Bekele, T.~M., Liu, H. (2017). Big scholarly data: A survey.
  IEEE Transactions on Big Data 3(1), 18--35.

\bibitem[{Xiao et~al.(2016)Xiao, Yan, Li, Jin, Wang, Yang, Chu, and
  Zha}]{xiao2016modeling}
Xiao, S., Yan, J., Li, C., Jin, B., Wang, X., Yang, X., Chu, S.~M., Zha, H.
  (2016). On modeling and predicting individual paper citation count over time.
  In: Proceedings of the Twenty-Fifth International Joint Conference on
  Artificial Intelligence (IJCAI-16). pp. 2676--2682.

\bibitem[{Yu et~al.(2014)Yu, Yu, Li, and Wang}]{yu2014citation}
Yu, T., Yu, G., Li, P.-Y., Wang, L. (2014). Citation impact prediction for
  scientific papers using stepwise regression analysis. Scientometrics 101(2),
  1233--1252.

\bibitem[{Zhang et~al.(2016)Zhang, Yu, Lu, Zhou, and Zhang}]{Zhang2016AdaWIRL}
Zhang, C., Yu, L., Lu, J., Zhou, T., Zhang, Z.~K. (2016). Adawirl: A novel
  bayesian ranking approach for personal big-hit paper prediction. In:
  International Conference on Web-Age Information Management. pp. 342--355.

\bibitem[{Zhang et~al.(2017)Zhang, Ning, Bai, Kong, Zhou, and
  Xia}]{Zhang2017Exploring}
Zhang, J., Ning, Z., Bai, X., Kong, X., Zhou, J., Xia, F. (2017). Exploring time
  factors in measuring the scientific impact of scholars. Scientometrics
  112(3), 1301--1321.

\bibitem[{Zhao et~al.(2015)Zhao, Erdogdu, He, Rajaraman, and
  Leskovec}]{Zhao2015SEISMIC}
Zhao, Q., Erdogdu, M.~A., He, H.~Y., Rajaraman, A., Leskovec, J. (2015).
  Seismic: A self-exciting point process model for predicting tweet popularity.
  In: ACM SIGKDD International Conference on Knowledge Discovery and Data
  Mining. pp. 1513--1522.

\end{thebibliography}

\end{document}